\documentclass[
10pt,
final,
twocolumn,
showpacs,
amsmath,
amssymb,
pra
]{revtex4}

\usepackage{graphicx}
\usepackage{dcolumn}
\usepackage{bm}

\newcommand{\hide}[1]{}
\newcommand{\eq}[1]{Eq.\,(\ref{#1})}

\def\+#1{\ifmmode{{#1}^{\dagger}}\else{$#1^{\dagger}$}\fi}
\def\*#1{\ifmmode{{#1}^{\ast}}\else{$#1^{\ast}$}\fi}
\def\bra#1{\ifmmode{\left< #1 \right|}\else{$\left< #1 \right|$}\fi}
\def\ket#1{\ifmmode{\left| #1 \right>}\else{$\left| #1 \right>$}\fi}
\def\braket#1#2{\ifmmode{\left< #1 | #2 \right>}\else{$\left< #1 | #2 \right>$}\fi}
\def\half{\ifmmode{\frac 1 2}\else{${1\over2}$}\fi}

\begin{document}
%
%
\title {
  Spectroscopic Temperature Determination of Degenerate Fermi Gases
  }
\author{
  Marijan Ko\v{s}trun and Robin C\^{o}t\'{e}
  }
\affiliation{
  Physics Department, University of Connecticut,
  2152 Hillside Rd., Storrs, Connecticut 06269-3046.
  }
\date{\today}

\begin{abstract}
  We suggest a simple method for measuring the temperature
  of ultra-cold gases made of fermions. We show that by using a 
  two-photon Raman probe, it is possible to obtain lineshapes
  which reveal properties of the
  degenerate sample, notably its temperature $T$. 
  The proposed method could be used with identical fermions in 
  different hyperfine states interacting via $s$-wave scattering or
  identical fermions in the same hyperfine state via $p$-wave 
  scattering. We illustrate the applicability of the method in 
  realistic conditions for $^6$Li prepared in two different hyperfine
  states. We find that temperatures down to 0.05 $T_{F}$ can be determined
  by this {\it in-situ} method.
\end{abstract}

\maketitle

\section{Introduction}

The experimental realization of Bose-Einstein condensation (BEC) 
in atomic samples \cite{Dalfovoetal1996} have stimulated
a new wave of theoretical and experimental studies of degenerate
systems in the dilute and weakly interacting regime. 
Several recent experiments have successfully reached the 
degenerate regime for ultra-cold atomic samples of fermions. 
For example,
temperatures of 0.5 $T_{F}$ or less, where $T_{F}$ is the 
Fermi temperature of the cold fermions, have been reported for
$^{40}$K~\cite{DeMarcoJin1999}, or
$^{6}$Li~\cite{Truscottetal2001:6li}. 
One of the goals of reaching even lower temperatures is the formation of
Cooper pairs of interacting fermions and the detection
of the Bardeen-Cooper-Schrieffer (BCS) phase transition in
a trapped gas of fermionic atoms. Such
transitions in atomic gases will allow the study of phenomena
usually associated with condensed matter physics, such as 
superconductivity. Recently, ``resonance superfluidity" 
\cite{holland} has been predicted in the strong-coupling 
regime occurring near a Feshbach resonance. This regime
is being explored experimentally \cite{thomas2002:science}.

One of the key parameters in studying these systems is the 
accurate value of the temperature $T$. Various methods
are being employed to determine $T$, such as
fitting the tail of the fermion energy distribution \cite{thomas:temperature},
using mixtures with bosons \cite{Truscottetal2001:6li}, 
or using impurities as probes \cite{Ferrari1999}.
Each of these approaches have limitations. For example, considering
that the thermal contribution in a Fermi gas vanishes with temperature,
fitting the tail of the energy distribution becomes very challenging. 
Even in a fermion-boson mixture with large attractive interactions,
such as $^{40}$K-$^{87}$Rb \cite{Ferrarietal2002:KRb,cond-mat/0301182},
or $^{6}$Li-$^{7}$Li~\cite{Truscottetal2001:6li},
it is questionable whether bosons and fermions stay in thermal equilibrium
when the temperature is lowered to degenerate regime.
Even if so, the determination of the 
bosons temperature is not easy below $T_{c}/2$ ($T_{c}$:
BEC critical temperature) \cite{modugno}. 
Typically, temperature is read by imaging the velocity distribution
of a freely expanded sample.

In this paper, we describe an {\em in situ} method to measure the
temperature of a sample based on spectroscopic measurements. 
It offers the potential for very 
accurate determination of extremely low temperatures in degenerate Fermi 
systems as well as other properties (e.g., their Fermi temperature 
$T_{F}$). Furthermore, if
used in systems made of fermions only, it avoids the problem arising
from the superfluidity of Bose-Einstein condensates when $T$ is
below the BEC critical temperature $T_{c}$, which suppresses 
scattering \cite{superfluidity}.
The approach is based on Doppler-free two-color spectroscopy. 
Such techniques have been successfully employed to create ultra-cold
dimers in ultra-cold bosonic gases and in 
BEC \cite{molecules}. 

\section{Method}

\begin{figure*}[htp]
  \includegraphics[scale=0.7,clip]{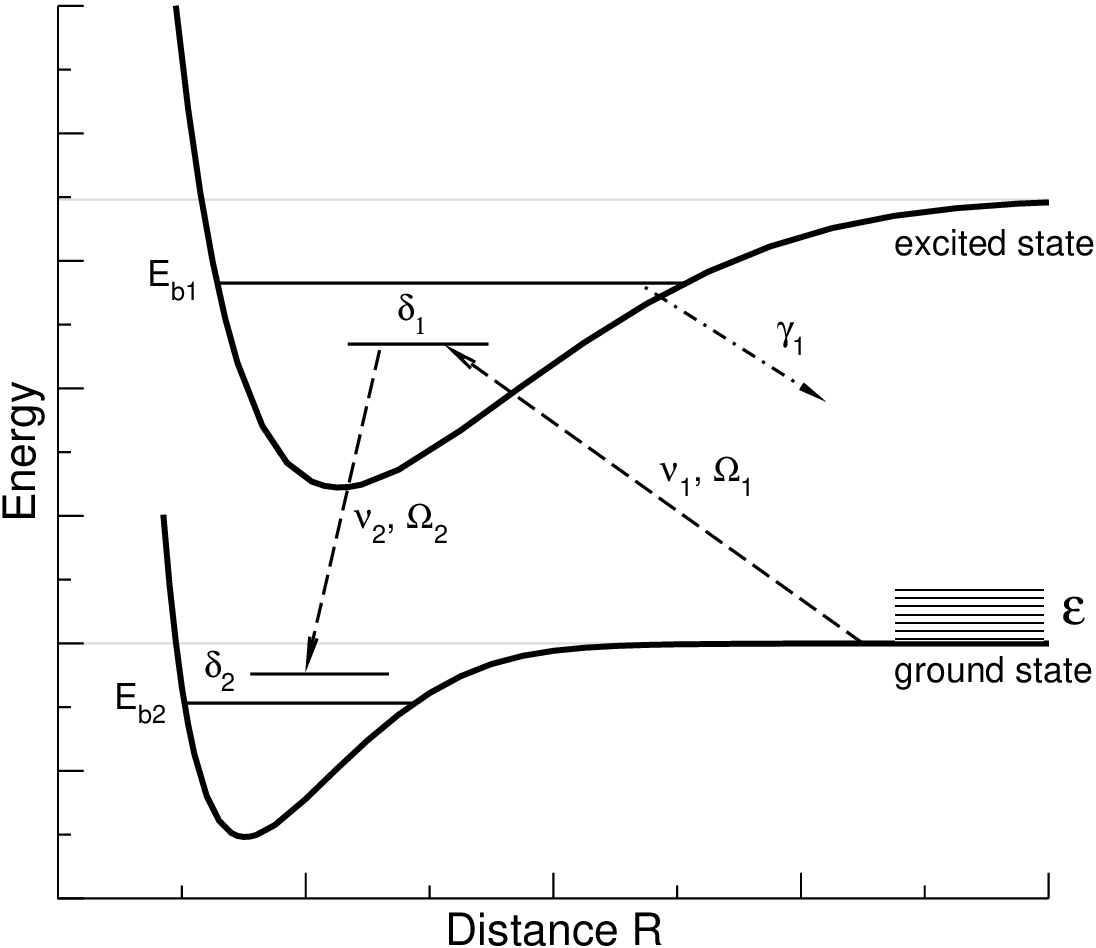}
  \caption{Two-photon scheme. In this setup the detunings are 
    $\delta_1 = E_{b1}-\nu_1 > 0$ and $\delta_2 = E_{b2}-\nu_1 +\nu_2 < 0$.
    }
  \label{fig1}
\end{figure*}

The schematic of the two-photon transition is illustrated in Fig.~\ref{fig1},
together with the relevant quantities. 
While two atoms
approach each other along the molecular ground state with a given 
relative kinetic energy $\epsilon$, two co-propagating lasers 
can induce a two-photon
transition into a bound level $E_{b2}$ of the molecular ground state.
The first laser of intensity $I_{1}$ is detuned
from the excited molecular bound level $E_{b1}$ by $\delta_1=E_{b1}-\nu_1$.
Similarly, the second laser of intensity $I_{2}$ is detuned 
by an amount  $\delta_{2}=E_{b2}-(\nu_1-\nu_2)$
from the bound level energy $E_{b2}$ of the ground molecular state.
The rate coefficient
for the photoassociation (PA) process is given by \cite{BohnJulienne1996}
\begin{equation}
   {\sf K}(T, L_{1}, L_{2})
   = \left\langle \frac{\pi v}{\kappa^{2}} \sum_l (2l+1)
     |S_{l}(\epsilon ,L_{1}, L_{2})|^{2}
     \right\rangle \; ,
   \label{eq:rate-def}
\end{equation}
where $L_{i}=\{ I_{i},\delta_{i} \}$ represent the parameters of laser $i$,
$\epsilon=\hbar^{2}\kappa^{2}/2\mu = \mu v^{2}/2$, $\mu$ is the 
reduced mass, and
$v$ is the relative velocity of the colliding pair. 
In \eq{eq:rate-def} the sum goes over contributing partial waves $l$, 
and $S_{l}$ represents the scattering matrix element
for producing the final state $E_{b2}$ from the initial continuum state.
Averaging over relative velocities is implied by
$\langle\dots\rangle$.
The scattering matrix is
well approximated by \cite{BohnJulienne1996}
\begin{equation}
  |S_{l}|^{2} = \frac{(\epsilon-\delta_{2})^{2} \gamma_{1}\gamma_{s}}
  {(\epsilon-\Delta_{+})^{2}(\epsilon-\Delta_{-})^{2} + 
  (\gamma^{2}/4)(\epsilon-\delta_{2})^{2}}
  \; ,
  \label{eq:S-def}
\end{equation}
where $\gamma_{1}$ is the width of the intermediate bound level $E_{b1}$ and 
$\gamma_{s}\simeq 4\pi^{2}(I_{1}/c)|\langle E_{b1}|D(R)|\epsilon,l\rangle|^{2}$
is the stimulated width from the continuum initial state $|\epsilon ,l\rangle$
to the intermediate state $|E_{b1}\rangle$. Here, $D(R)$ is the molecular
dipole transition moment, and in the zero-energy limit, 
$\gamma_{s}\propto \epsilon^{1/2+l}$. Also, 
$\gamma=\gamma_{1}+\gamma_{s} \simeq\gamma_{1}$ if the laser intensities 
are not too large (this is the regime we are
interested in). Finally, $\Delta_{\pm}$ represent a split
in the single resonance due to the second laser, and is given by
\begin{equation}
   \Delta_{\pm} = \frac{1}{2}(\delta_{1}+\delta_{2}) \pm \frac{1}{2}
   \sqrt{(\delta_{1}-\delta_{2})^{2}+4h^{2}\Omega_{2}^{2}} \; ,
   \label{eq:delta-pm}
\end{equation}
where $h^{2}\Omega_{2}^{2}=(2\pi I_{2}/c)|\langle E_{b1}|
D(R)|E_{b2}\rangle|^{2}$ defines the Rabi frequency $\Omega_{2}$.
From the analytical form of $|S_{l}|^{2}$, we expect two peaks located
at $\epsilon\simeq\Delta_{\pm}$, and a minimum ($|S_{l}|^{2}=0$) 
located between
the peaks at $\epsilon=\delta_{2}$ (due to
destructive interference between the two scattering paths).  
Because of the $\epsilon^{1/2+l}$ Wigner's 
threshold behavior of $\gamma_{s}$, the peak at $\Delta_{-}$ will 
be weaker than the one at $\Delta_{+}$.

We consider samples of identical fermions of mass $m$ in the ultra-cold 
regime (so that $2\mu =m$). While
there is no $s$-wave ($l=0$) scattering 
between two identical fermionic atoms in the same hyperfine state 
(the lowest contribution is then a $p$-wave ($l=1$) which vanishes 
as $\epsilon\rightarrow 0$), $s$-wave scattering
occurs between different hyperfine states.
We will concentrate our treatment to the latter case, and assume
two identical fermions in hyperfine states labeled 
1 and 2, respectively.

To compute the rate coefficient ${\sf K}$, we need to average over
the distribution of relative velocities in the system.
The velocity distribution of each ensemble follows a Fermi 
distribution characterized by its temperature $T$ and
Fermi energy $E_{F}=k_{B}T_{F}$. Because of the setups in 
most experiments, we consider the atoms trapped in an harmonic
potential. For large traps with frequencies $\omega_{1},\omega_{2}$, 
and $\omega_{3}$,
the density of state for the hyperfine state $i$ is given by
$\rho(\vec{k}_{i})=\hbar^{3}k_{i}^{3}/m^{3}\omega_{1}\omega_{2}\omega_{3}$.

The Fermi energy $E_{F}$ of the atoms in a particular hyperfine state 
is the energy of the highest occupied state in the harmonic potential
at absolute zero and so is determined by the total number of atoms sharing
the same hyperfine state.
To have the same number of atoms in each hyperfine state provides two
benefits: both group of atoms have the same Fermi energy, and the s-wave
scattering rate is maximal.
This requirement in not too stringent experimentally:
if the sample is to be prepared from the atoms in a single hyperfine
state by transferring $\sim50\%$ of the atoms to another hyperfine state,
then the maximal number difference between the two
is determined by the degeneracy of the trap potential at the Fermi energy.
E.g., for isotropic harmonic trap with the degeneracy of the $m$-th energy
level $\sim m^2$ it follows $\Delta N/N \approx \cdot N^{-1/3}$.
In what follows we assume that both hyperfine states have identical
number of atoms so that their Fermi energies are identical as well.

\begin{figure*}[htp]  
  \includegraphics[scale=0.6,clip]{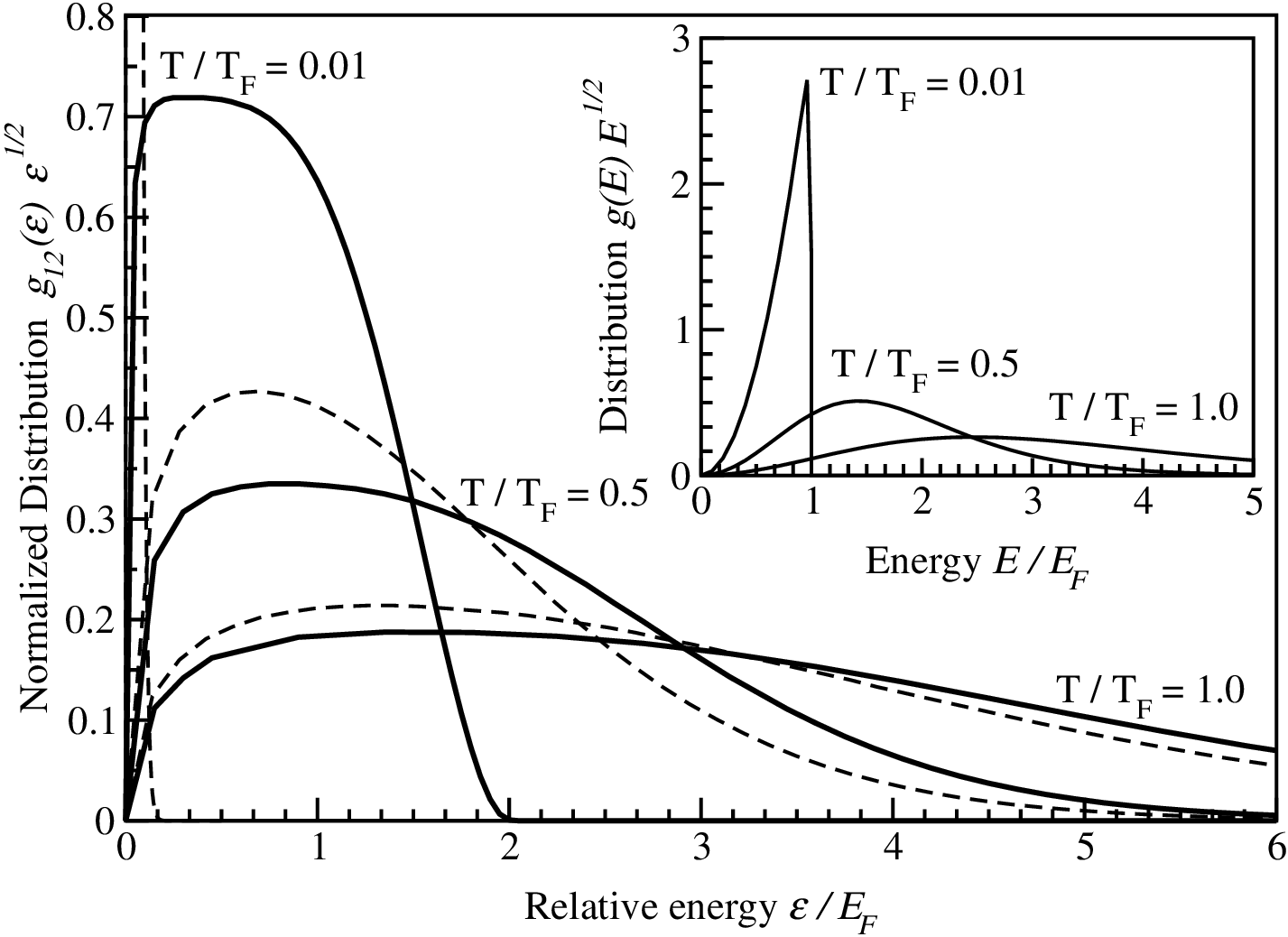}
  \caption{
    Relative velocity distribution for various temperatures:
    $g_{12}(\vec{\kappa})d^{3}\kappa = 2\pi (2\mu/\hbar^{2})^{3/2}
    g_{12}(\epsilon) \sqrt{\epsilon}d\epsilon$ (here,
    $g_{12}(\epsilon)\sqrt{\epsilon}$ is
    normalized to unity). 
    The effect of degeneracy becomes apparent
    when $g_{12}$ (solid lines) are compared to the 
    corresponding 
    Maxwell-Boltzmann distributions (dashed lines). 
    The inset shows the corresponding single-particle 
    distribution $g(E)\sqrt{E}$.
    }
  \label{fig2}
\end{figure*}

The distribution of the relative momentum $\vec \kappa$ (normalized to
unity) is 
\begin{eqnarray}
  \label{eq:frel}
   g_{12}(\kappa)& = & \int d^{3}k_{1}\; d^{3}k_{2}\;  
   g(\vec{k}_{1}) \; g(\vec{k}_{2})\; 
   \delta (\vec{k}_{1}-\vec{k}_{2}-2\vec{\kappa})  \;, \nonumber \\
   & = & \int d^{3}k_{2} \; g(\vec{k}_{2})\;
    g(\vec{k}_{2}+2\vec{\kappa}) \; ,
\end{eqnarray}
where $g(\vec{k}_{i})\equiv \rho(\vec{k}_{i})
f_{\rm FD}(\vec{k}_{i},E_{F},T)/N_{i}$. Here, the function
$f_{\rm FD}(\vec{k}_{i},E_{F},T)=
\{ 1+\exp[(E_{{\vec k}_{i}}-E_{F})/k_{B}T] \}^{-1}$ is
the Fermi-Dirac distribution with
$E_{{\vec k}_{i}} = \hbar^2 k_{i}^2/2 m$, and 
$N_{i}=\int d^3 k_{i} \rho(\vec{k}_{i})f_{\rm FD}(\vec{k}_{i},E_{F},T)$
is the number of fermions in the hyperfine state $i$. 
Note that $g_{12}(\kappa)$ 
depends on $E_{F}$ and $T$, but not on the direction of $\vec{\kappa}$:
the integration removes that dependence. 
In Fig.~\ref{fig2}, we illustrate the distribution
$g_{12}$ in term of the 
relative energy $\epsilon$ for various temperatures and compare
it with the corresponding Maxwell-Boltzmann (MB) distributions.
As the temperature drops under $T_{F}$, the difference between
the two types of distribution becomes more apparent; as the
system becomes more degenerate, the $g_{12}$ distribution 
spreads over a larger range of energies (when compared
to MB distributions). As $T$ approaches zero, $g_{12}$
is entirely contained between $\epsilon=0$ and $2E_{F}$. As
opposed to the single particle distribution $g(E)$ (see inset),
which is contained between $E=0$ and $E_{F}$ at zero-$T$
and decreases sharply to zero at $E_{F}$, $g_{12}$ goes to zero
more gradually. In fact, for the case of free particles, one
can show that $g_{12}(\epsilon)$ is a triangle with values 1 at 
$\epsilon =0$ and 0 at $\epsilon = 2 E_{F}$, while $g(E)$
is the standard step function between $E=0$ and $E_{F}$.

We want to use 2-photon scattering to probe the relative
velocity (or energy) distribution and infer the temperature
of the system. 
At ultra-cold temperatures, 
Eq.(\ref{eq:rate-def}) becomes 
${\sf K}=(\pi\hbar/2m)\int d^{3}\kappa\; g_{12}(\kappa) |S_{0}|^{2}/\kappa$,
which can be rewritten in terms of $\epsilon$ as
\begin{equation}
  \label{eq:ratewrenergy}
  {\sf K}(T, L_{1}, L_{2}) = \frac{4\pi^{2}}{\hbar}
  \int_{0}^{\infty} d\epsilon \, g_{12}(\epsilon)
  |S_{0}(\epsilon ,L_{1}, L_{2})|^2 \; .
\end{equation}
Besides $T$ and $E_{F}$, ${\sf K}$
depends on the detunings $\delta_1$ and $\delta_2$ and the Rabi frequencies
$\Omega_1$ and $\Omega_2$. Because of the important variation of
$|S_{0}|^{2}$ with respect to $\delta_{2}$, lineshapes obtained as a
function of $\delta_{2}$ contain precious information about
the degenerate Fermi gas. As $T$ is reduced, the distribution $g_{12}$
becomes more sharply defined for $\epsilon$ between 0 and $2E_{F}$, 
and by scanning $\delta_{2}$ for nearby values, the large variations
in the integrand of Eq.(\ref{eq:ratewrenergy}) will probe $g_{12}$.

We computed lineshapes for realistic systems.
The Fermi energies $E_{F}$ reached in todays experiments are in the range
of few $\mu$K. In addition to this experimental constraint, one needs to
be detuned far enough from resonance to not populate the level $v_{1}$,
with typical detunings $\delta_{1}\sim 150$ MHz 
(7.2 mK) \cite{randy-comment}. The lifetime of most levels $v_{1}$ is
about 10-20 nsec, giving $\gamma_{1}\sim 16-32$ MHz (0.76-0.38 mK)
Finally, one needs
intensities large enough to drive the transition, but not too large
to heat the system (through broadening). Typical values for high lying
states $v_{1}$ at 500 GHz below the asymptote are $\Omega_1 = 100$ kHz
(4.8 $\mu$K) and $\Omega_{2}\sim 2$-3 MHz (96-144 $\mu$K). 
Under these conditions, we have
$\delta_{1}\gg h\Omega_{2}$. With large $\delta_{1}$, $\Delta_{+}$
will also be large, and the only way to get a peak of $|S|^{2}$
with $\epsilon \sim E_{F}$ will come from $\Delta_{-}$: hence, we must
have $\delta_{2}\sim E_{F}$ as well. Then, 
$\delta_{1}\gg h\Omega_{2}\gg\delta_{2}$ and we will have
peaks at $\Delta_{-}\sim \delta_{2} - h^{2}\Omega_{2}^{2}/\delta_{1}$
and $\Delta_{+}\sim \delta_{1} + h^{2}\Omega_{2}^{2}/\delta_{1}$, 
as well as a minimum at $\epsilon =\delta_{2}$. 

\section{Results and Discussion}

\begin{figure*}[htp]  
  \includegraphics[scale=0.5,clip]{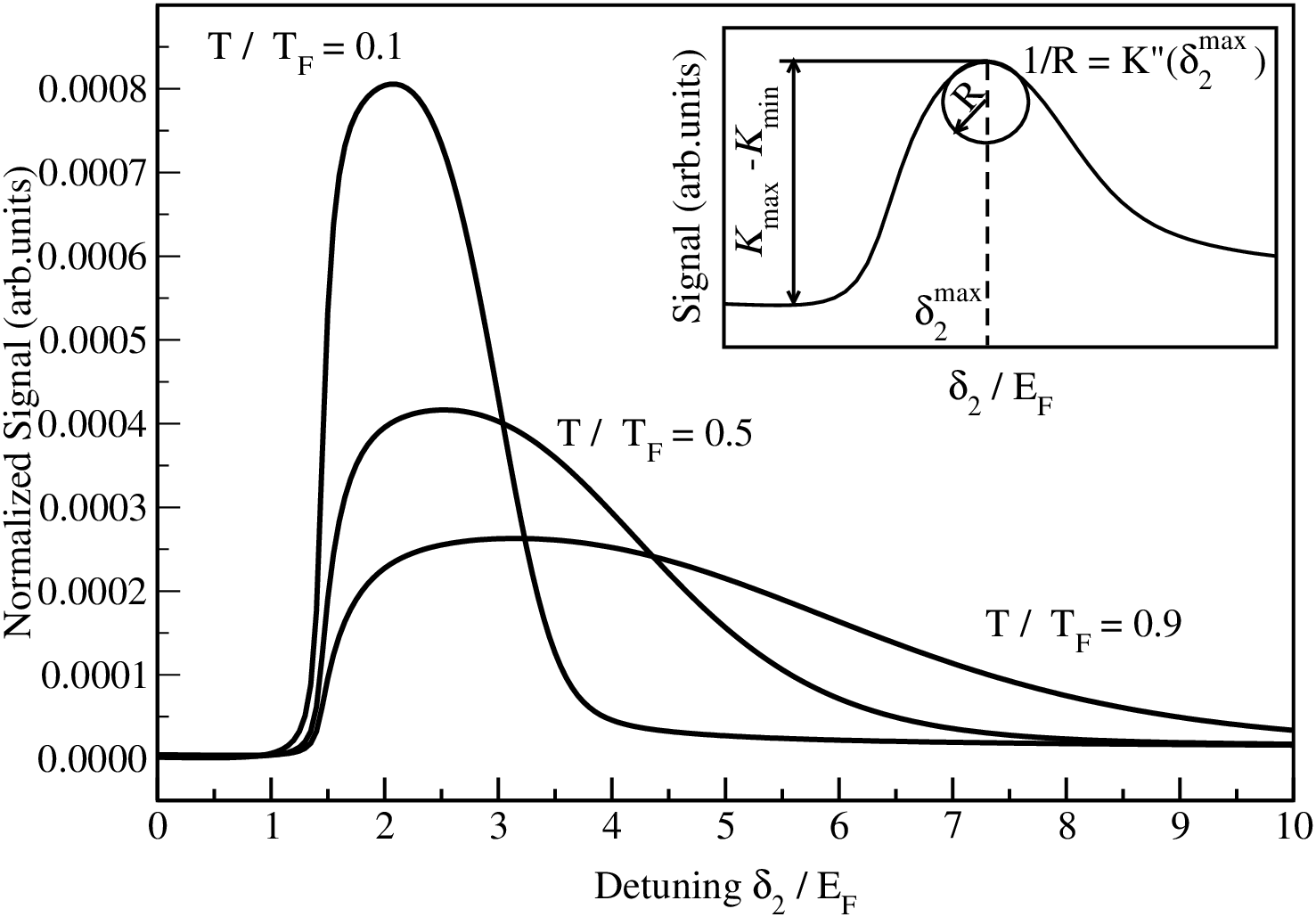}
  \caption{
    Lineshapes as the functions of $\delta_2$ for temperatures
    $T/T_{F} = 0.1$, 0.5, and 0.9, with $\gamma_{1}=3$ mK, $\delta_{1}=0.56$ mK,
    and $\Omega_{2}=100$ $\mu$K. The definitions of ${\sf K}''$,
    ${\sf K}_{\rm max}$, ${\sf K}_{\rm min}$, and $\delta_{2}^{\rm max}$
    are illustrated in the inset.
    }
  \label{fig3}
\end{figure*}

We obtained lineshapes for various temperatures by varying 
$\delta_{2}$ in (\ref{eq:ratewrenergy}). We selected parameters
consistent with the case of $^{6}$Li (although similar
results can be obtained for other species): $E_{F} = 1$ $\mu$K, 
$\delta_{1}= 7$ mK, $\gamma_1 = 0.76$ mK 
and $h\Omega_{2}= 100$ $\mu$K. The remaining 
parameters lead to a multiplicative constant in ${\sf K}$ that can be 
factored out for all lineshapes. We show representative lineshapes
normalized to unity (in arbitrary units) in Fig.~\ref{fig3};
as $T$ is lowered below $T_{F}$, the peak becomes narrower and sharper.
Many other features can be identified on those lineshapes. 
As $\delta_{2}$ grows, ${\sf K}$ reaches a minimum value ${\sf K}_{\rm min}$, 
increases rapidly to its maximum value ${\sf K}_{\rm max}$, and then decreases
more slowly to a ``background" value ${\sf K}_{\rm bg}$. 
In addition, inflection points on both sides of the peak can be identified.
All these features can be extracted from the lineshapes, and could be 
used to extract temperature. 
For example, one could take the ratio 
$[{\sf K}_{\rm max}-{\sf K}_{\rm bg}]/[{\sf K}_{\rm min}-{\sf K}_{\rm bg}]$ 
of a given lineshape, in principle removing the multiplicative constant
in ${\sf K}$. 
However, this ratio varies very little for $T/T_{F}\sim 0.5$ 
and below. 
Besides the effect of a background signal/noise would also affect/reduce
the sensitivity of the temperature determination. 
One could also map the location (i.e. the value of $\delta_{2}$) 
of these features as a function of $T$, but again, 
their variations are small below say $T/T_{F}\sim 0.1$. 

Alternative features, such as ratios of the derivatives at the inflection
points (on each side of the main peak) could be used. In this case, the
background would not play any role, nor the multiplicative constant
in ${\sf K}$. 
A simple measure incorporating both criteria is the ratio
of the curvature of the signal at the peak value and its height as defined
from the minimum value of ${\sf K}$ (see inset in Fig.~\ref{fig3}), i.e.
\begin{equation}
   \label{eq:ratio}
   C \equiv -\frac{{\sf K}''(\delta_{2}^{\rm max})}
                   {{\sf K}_{\rm max}-{\sf K}_{\rm min}} E_{F}^{2}\; .
\end{equation}
While finding the second derivative of a lineshape with respect to
$\delta_2$ may be challanging in general, at the local peak maximum
this should not be a problem as the first derivative there is zero.

\begin{figure*}[htp]  
  \includegraphics[scale=0.5,clip]{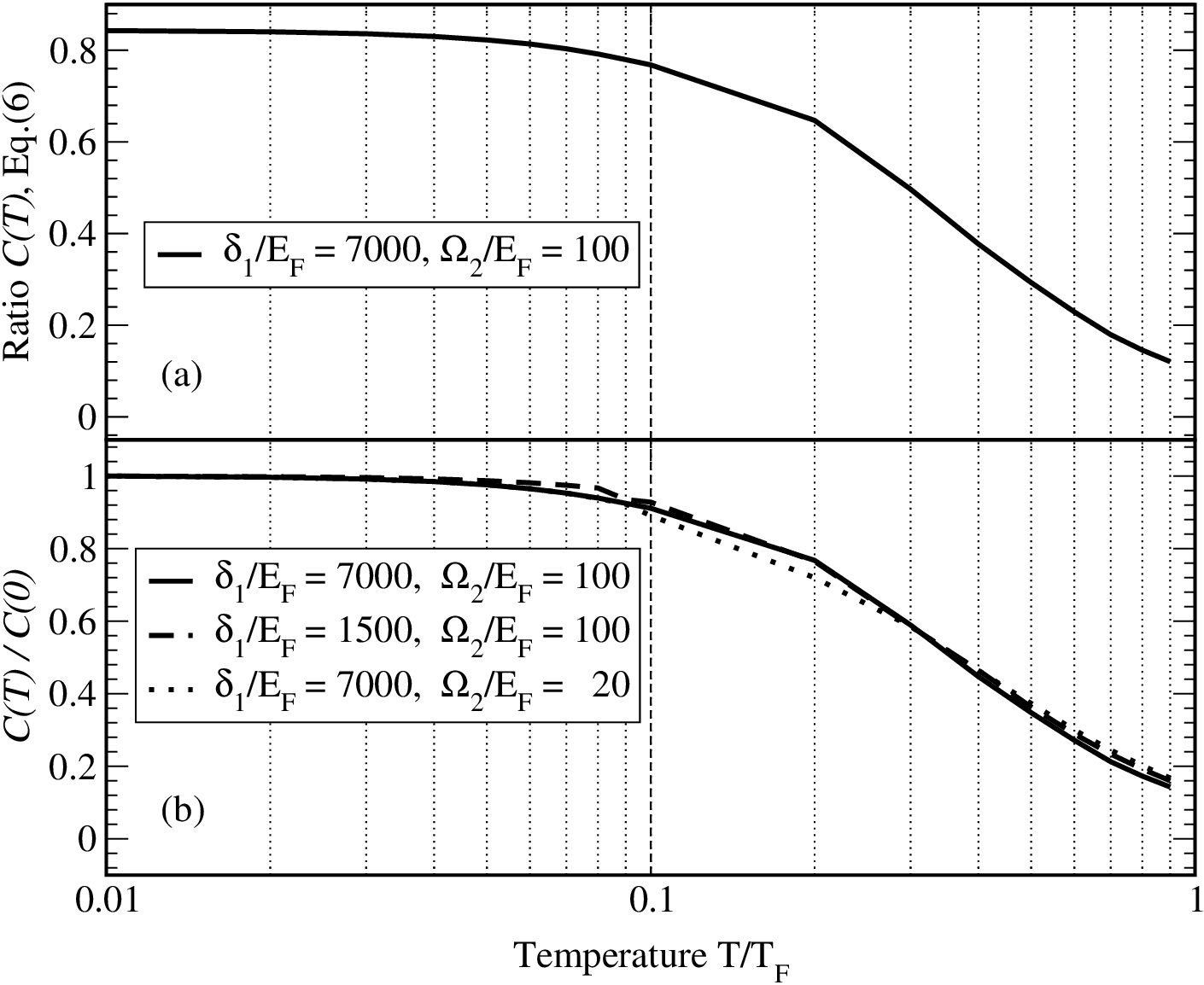}
  \caption{Ratio $C$ from Eq.(\ref{eq:ratio}) as a function 
           of $T/T_{F}$. The sensitivity of $C$ on other
           parameters is also illustrated by two other cases
           (dashed lines).}
  \label{fig4}
\end{figure*}

Because of the rapid increase of ${\sf K}$ from its minimum value to
its peak value, only a small range of $\delta_{2}$ needs to be scanned
(as compared to the ratio of the derivatives at the inflection points):
for $E_{F}\sim 1$ $\mu$K, scans with kHz precision are required.
In Fig.~\ref{fig4}, we show $C$ as a function of $T/T_{F}$: it decreases
monotonically with $T$, and even at $T/T_{F}\sim 0.05$, its variation
is noticeable. So, by measuring three features on the lineshape, 
namely its minimum, maximum, and the curvature at the maximum, one
can evaluate $C$ and determine the temperature of the degenerate
Fermi sample by reading it from Fig.~\ref{fig4}. Although the exact 
values of $C$ depend on other parameters ($\gamma_{1}$, $\delta_{1}$,
and $\Omega_{2}$), its general shape follows the example shown
in Fig.~\ref{fig4}. For illustration purposes, we also include
two other curves (corresponding to other sets of $\delta_{1}$
and $\Omega_{2}$). The curve with the largest variations at
low $T/T_{F}$ is found when we have 
$\delta_{1}\gg h\Omega_{2}\gg\delta_{2}$ with a factor of 100
between each parameter.

The lineshapes shown in Fig.~\ref{fig3} are not very sensitive 
to small variations in $\gamma_{1}$, $\delta_{1}$ or $\Omega_{2}$:
variations of few \% give very similar curves. Of these parameters,
$\gamma_{1}$ cannot be varied experimentally, except by changing
the intermediate molecular level $E_{b1}$. However, for most atoms with
Fermi species considered in todays experiments, i.e., alkali metals,
$\gamma_{1}$ varies little with $E_{b1}$ (e.g., for $^{6}$Li$_{2}$
in the triplet excited state, $\gamma_{1}$ varies from 
$\sim 11.7$ MHz down to 3.2 MHz (or 0.56 mK to 0.16 mK). 
However, systems with even smaller $\gamma_{1}$ (about 1-10 $\mu$K)
would offer
the possibility of measuring the velocity distribution $g_{12}$
directly by scanning $\delta_{2}$. In fact, $|S|^{2}$ then becomes
proportional to the delta-function $\delta (\epsilon -\delta_{2})$.
The extremely small variations in $g_{12}$ as $T\rightarrow 0$
could possibly be detected. Another possibility to enhance the
sensitivity of this type of measurement could be realized using 
Feshbach resonances: a sharp increase in the integrand of 
Eq.(\ref{eq:ratewrenergy}) at $\epsilon\sim \epsilon_{\rm res}$,
where $\epsilon_{\rm res}$ is the resonant relative energy,
would emphasize the corresponding region of the velocity distribution.
By positioning the resonance appropriately (e.g., via external magnetic 
fields), strong contribution of regions of $g_{12}$ where small 
variations occur as $T\rightarrow 0$ could be obtained, 
hence more precise information could be extracted from
the lineshapes. Notice that $p$-wave Feshbach
resonances \cite{p-wave} could be used to enhance $p$-wave scattering in a
single-fermion population. 

\section{Conclusion}

We have shown that the lineshapes obtained using
a two-photon Raman probe can yield very precise determination
of the temperature in degenerate Fermi gases. One of the
key advantage of this approach, beside its 
non-destructive nature, is the extremely low temperatures
measurable: in the example considered here, as low as $\sim 0.05 T_{F}$. 
Although we have illustrated this two-color scheme
for identical fermions in two different internal states
with equal populations, this method can be made much more flexible.
For example, if the populations are very different, the $g_{12}$
distribution may exhibit sharper structures 
that could be used 
advantageously. The enhancement due to 
Feshbach resonances could also help probing chosen regions of 
$g_{12}$, and extend the method to samples of identical fermion
in a single internal state, by using $p$-wave Feshbach resonances.
Finally, when the interaction between the fermions is attractive
such as for $^6$Li, Cooper pairs may form. 
Photoassociation of a Cooper pair should demand less energy than the
photoassociation of the two fermions produced by breaking either one
or two Cooper pairs \cite{TormaZoller2000}.
These processes should contribute additional features
on top of the simple lineshapes, their locations and magnitudes
probing the energy gap and the number of Cooper pairs.

We thank R.G. Hulet and M. Modugno for helpful discussions. This
work was supported by the National Science Foundation Grant
PHY-0140290 and by the University of Connecticut Research 
Foundation.

%
%

\end{document}